\documentstyle[aps]{revtex}
\draft
\twocolumn

\title{Correlation functions for
time-dependent calculation of linear-response functions}

\author{ Toshiaki Iitaka \\
Nanoelectronics Materials Group, \\
Frontier Research Program, RIKEN, \\
Hirosawa 2-1, Wako, Saitama 351-01, Japan 
}

\begin{document}
\date{(Received 19 August 1997)}

\maketitle



\begin{abstract}
We emphasize the importance of choosing an appropriate correlation function to reduce numerical errors in calculating the linear-response function as a Fourier transformation of a time-dependent correlation function.
As an example we take dielectric functions of silicon crystal calculated with a time-dependent method proposed by Iitaka {\it et al.} [Phys. Rev. E {\bf 56}, 1222 (1997)].
\end{abstract}

\begin{flushleft}
\end{flushleft}

\narrowtext

Recently, Iitaka {it et al.} proposed a linear scaling algorithm for calculating linear-response functions of noninteracting electrons \cite{Iitaka97}.
This algorithm has been successfully applied to the calculation of dielectric functions of Si crystal \cite{Iitaka97} and Si nanocrystallites \cite{Nomura97}.
They calculate the linear-response of $B$ due to the perturbation $A e^{-i\omega t}$ as
\begin{eqnarray}
\label{eq:chi.time.one}
\chi_{BA}(\omega+i\eta) 
&=& 
\label{eq:chi.numerical.1}
\left\langle \!\!\! \left\langle
\rule{0pt}{24pt}
\int_0^T \!\!\!dt \ \ e^{+i(\omega + i\eta)t}  K(t)
\right\rangle \!\!\! \right\rangle
,
\end{eqnarray}
where the double angular brackets indicate the statistical average over random vectors $| \Phi \rangle $ and $K(t)$ is the time-dependent correlation function defined by
\begin{equation}
\label{eq:correlation}
K(t)= 2 \ {\rm Im}
\langle \Phi |  \theta(E_f-H)e^{+iHt}B e^{-iHt} \theta(H-E_f)A  | \Phi \rangle 
.
\end{equation}
In the case of the dielectric function $\epsilon_{\beta,\alpha}=1+4\pi \chi_{\beta \alpha}$ with $\alpha,\beta=x,y,z$ the perturbation and the response are the external electric field and polarization, respectively,
\begin{eqnarray}
\label{eq:dipole.operator1}
\langle j |A|i \rangle  &=& \langle j | x_{\alpha} |i \rangle ,
 \\
\label{eq:dipole.operator2}
\langle i|B |j \rangle &=&  \frac{-1}{V} \langle i  |x_{\beta}|j \rangle ,
\end{eqnarray}
where $V=L^3$ is the volume of the supercell.

In Ref.\cite{Iitaka97} Iitaka {\it et al.} further modified the dipole moment operator $x$ to the current operator $j$ by partial integration
\begin{eqnarray}
\label{eq:chi.dielectric.numerical}
\chi_{\beta \alpha}(\omega+i\eta) 
&=& 
\left\langle \!\!\! \left\langle
\rule{0pt}{24pt}
\int_0^T \!\!\!dt \ \ e^{-\eta t} 
\frac{e^{+i \omega t}-\delta_{\beta \alpha} }
{(\omega+i \eta)^2}
K_j(t)
\rule{0pt}{24pt}
\right\rangle \!\!\! \right\rangle , \\
\label{eq:chi.correlator.numerical}
K_j(t)&=&
\frac{-2}{V } {\rm Im}
\langle \Phi |  \theta(E_f-H)e^{+iHt} j_{\beta} e^{-iHt} \times \nonumber \\
&&     \theta(E_{cut}-H) \theta(H-E_f) j_{\alpha}  | \Phi \rangle 
.
\end{eqnarray}
and calculated the dielectric function 
 of silicon crystal consisting of $2^{15}$ Si atoms in a cubic supercell of $16^3$ unit cells with energy resolution $\eta=0.05 \ {\rm (eV)}$. Each unit cell is divided into $8^3$ cubic meshes. Iitaka {\it et al.} \cite{Iitaka97} used the empirical local pseudopotential developed by by Wang and Zunger in Ref.~\cite{Zunger} and one random vector for the statistical average. Figure~\ref{fig:silicon.crystal.eps} shows the result calculated with the same condition as in Ref.\cite{Iitaka97}.
Iitaka {\it et al.} attributed the divergence  of calculated response functions at low frequencies to the artificially introduced finite imaginary part \( \eta \) of the frequency.

However, closer examination suggests that the divergence may originate from the modification of  the position-position correlation function to the current-current correlation function. Assuming that the dielectric function is analytic at zero frequency, we easily notice that the low-frequency component of $K_j(t)$ has a frequency dependence of $\omega^{2-\delta_{\beta \alpha}}$. Therefore, numerical results calculated with $K_j(t)$ may easily suffer roundoff errors when the small amplitude of the low-frequency components is extracted from the large background of higher-frequency components. 

To check this scenario, we calculate the dielectric function without modification of dipole operators to current operators. Instead we modified Eqs. (\ref{eq:dipole.operator1}) and (\ref{eq:dipole.operator2}) to a finite-wave-number form to conform with periodic boundary conditions
\begin{eqnarray}
\langle j |A|i \rangle  &=& \left\langle j \left| \frac{e^{i q x_{\alpha}}}{i q} \right|i \right\rangle ,
 \\
\langle i|B |j \rangle &=&  \frac{-1}{V} \left\langle i  \left|\frac{e^{-i q x_{\beta}}}{-i q} \right| j \right\rangle ,
\end{eqnarray}
where $q$ is the smallest wave number in the supercell $q=2\pi/L$. 
The results are shown in Fig.~\ref{fig2}, where
other conditions are the same as in Fig.~\ref{fig:silicon.crystal.eps}.
 Here we obtained the dielectric function without any divergence at low frequencies, while the fluctuations at high frequencies are larger. We think that this result confirms our scenario that the divergence originates from the use of $K_j(t)$ instead of $K(t)$.

In summary, we found the importance of choosing an appropriate form of the correlation functions in computing linear-response functions as a Fourier transformation of time-dependent correlation functions. By choosing the current-current correlation function for high frequencies and the position-position correlation function for low frequencies, we may accurately calculate linear-response functions at any frequency.

\begin{figure}
\caption{
$\epsilon_{xx}(\omega)$ of silicon crystal calculated with current-current correlation function $K_j(t)$. The system consists of $2^{15}$ Si atoms in a cubic supercell of $16^3$ unit cells. Each unit cell is divided into $8^3$ cubic meshes. The energy resolution is $\eta= 0.05 {\rm (eV)}$. We used the empirical local pseudopotential in Ref.~\protect\cite{Zunger}.
(a) Real part and (b) imaginary part.
}
\label{fig:silicon.crystal.eps}
\end{figure}

\begin{figure}
\caption{
Same as Fig.~\protect\ref{fig:silicon.crystal.eps} but
calculated with position-position correlation function $K(t)$. 
}
\label{fig2}
\end{figure}


\begin{thebibliography}{99}

\bibitem{Iitaka97} T.~Iitaka, S.~Nomura, H.~Hirayama, X.~Zhao, Y.~Aoyagi and T.~Sugano, Phys. Rev. E {\bf 56}, 1222 (1997).

\bibitem{Nomura97}
S.~Nomura,  T.~Iitaka, X.~Zhao, T.~Sugano, and  Y.~Aoyagi, Phys. Rev. B {\bf 56} 4348 (1997).

\bibitem{Zunger}
L.W.~Wang and A.~Zunger, J.~Chem. Phys. {\bf 100}, 2394 (1994).

\end{thebibliography}
\end{document}